\documentclass[times, twoside]{zHenriquesLab-StyleBioRxiv}
\leadauthor{Kaya}

\renewcommand{\bBioRXiv}{{\footnotesize\footerfont \thepage}}
\renewcommand{\cBioRXiv}{{\footnotesize\footerfont \thepage}}
\renewcommand{\dBioRXiv}{{\footnotesize\footerfont \thepage}}

\begin{document}
\title{Differentiating Physical and Psychological Stress Using Wearable 
Physiological Signals and Salivary Cortisol}
\shorttitle{Wearable Stress Differentiation with Salivary Cortisol}
\author[1]{Ozan Kaya}
\author[1]{Nikoletta Athanassopoulou}
\author[2]{George G. Malliaras}
\author[2,3,4]{Marco Vinicio Alban-Paccha}
\affil[1]{Institute for Manufacturing, Department of Engineering, 
University of Cambridge, Cambridge, CB3 0FS, UK}
\affil[2]{Electrical Engineering Division, Department of Engineering, 
University of Cambridge, Cambridge, CB3 0FA, UK}
\affil[3]{School of Electrical and Electronic Engineering, 
University College Dublin, Dublin, Ireland}
\affil[4]{Division of Anaesthesia, Department of Medicine, 
University of Cambridge, Addenbrooke's Hospital, Cambridge, CB2 2QQ, UK}
\maketitle

\begin{abstract}
\textbf{Objective:} This study aimed to assess how wearable physiological signals, 
alone and combined with salivary cortisol, distinguish physical and psychological 
stress and their recovery states. 
\textbf{Methods:} Six healthy adults completed three laboratory sessions on separate 
days: rest, physical stress (high-intensity cycling), or psychological stress 
(modified Trier Social Stress Test). Heart rate, heart rate variability, 
electrodermal activity, and wrist accelerometry were recorded continuously, and 
salivary cortisol was sampled at five time points. Features were extracted in 
non-overlapping 10-minute windows and labelled as rest, physical stress, physical 
recovery, psychological stress, or psychological recovery. A gradient boosting 
classifier was trained using wearable features alone and with five additional 
cortisol features per window. Performance was evaluated using 
leave-one-participant-out cross-validation.
\textbf{Results:} Wearable-only classification achieved 77.8\% overall accuracy, 
with high accuracy for physical stress and recovery but frequent misclassification 
of psychological stress and recovery (recall 50.0\% and 54.2\%). Including cortisol 
improved overall accuracy (94.4\%), particularly for psychological states, increasing 
recall to 83.3\% and 87.5\%. Cortisol also reduced misclassification between 
psychological stress and rest.
\textbf{Conclusion:} Wearable signals alone were insufficient to reliably distinguish 
psychological stress from rest and recovery. Integrating salivary cortisol improved 
classification of psychological stress and recovery and reduced confusion with rest, 
highlighting the value of endocrine context alongside wearable physiology.
\textbf{Significance:} These findings support multimodal stress monitoring and 
motivate larger, ecologically valid studies and scalable alternatives to repeated 
cortisol sampling.
\end{abstract}


\begin{corrauthor}
marco.albanpaccha\at ucd.ie
\end{corrauthor}

\section*{Introduction}
Stress is a dynamic process involving coordinated psychological and biological 
responses to perceived challenges and demands \cite{b1,b2}. Prolonged or 
dysregulated stress responses contribute to adverse mental and physical health 
outcomes. This has led to increasing interest in monitoring stress as it unfolds 
in daily life rather than relying solely on retrospective assessments \cite{b3,b4,b5}. 
Beyond determining whether stress is present, meaningful interpretation of stress 
responses requires understanding the context and source of physiological activation, 
particularly before clinically significant symptoms emerge.

Traditional approaches to stress assessment rely primarily on self-report 
questionnaires or on controlled laboratory stress induction paradigms such as the 
Trier Social Stress Test (TSST) \cite{b6,b7,b8}. These approaches have generated 
rich insights into stress reactivity, but they are limited by recall bias, 
participant burden, and restricted ecological validity \cite{b9,b10}. These 
limitations have driven interest in passive sensing methods that can capture 
stress-related physiological changes without requiring continuous active input from 
individuals \cite{b11,b12}.

Wearable devices can continuously measure signals such as heart rate, heart rate 
variability (HRV), electrodermal activity (EDA), and movement, and are widely used 
in mobile health applications for stress detection \cite{b13,b14,b15,b16,b17}. 
Under controlled conditions, these autonomic markers exhibit consistent changes in 
response to stressors \cite{b18,b19}. However, autonomic signals primarily index 
general arousal and are strongly influenced by physical activity, posture, emotional 
valence, and environmental context \cite{b20,b21,b22}. Consequently, wearable-based 
stress detection models tend to conflate physiologically distinct states that share 
similar arousal levels and often struggle to separate psychologically induced stress 
from physical exertion, recovery from exercise, or quiet rest \cite{b23,b24,b25,b26,b27}. 
This lack of specificity limits the interpretability of wearable-derived ``stress 
scores,'' especially in mental health applications where knowledge of psychological 
stress is central \cite{b28}.

In addition to autonomic activation, stress involves hormonal processes that reflect 
the appraisal of stressors and engagement of the hypothalamic--pituitary--adrenal 
(HPA) axis. Salivary cortisol is a well-established marker of HPA axis activity and 
is more tightly linked to psychological stress exposure than to short-lived changes 
in heart rate or skin conductance \cite{b29,b30,b31}. Cortisol responses classically 
exhibit a delay and a prolonged time course relative to fast autonomic changes, 
suggesting that they may provide complementary temporal and contextual information 
about stress that is not captured by wearable signals alone \cite{b32,b33}. Despite 
this, cortisol is rarely integrated directly with wearable data in stress 
classification models, and its value for differentiating stress types in a multimodal 
setting remains underexplored \cite{b34}.

This proof-of-concept study addresses this gap by combining wearable physiological 
signals with salivary cortisol in a small, tightly controlled experiment. We induced 
physical stress using high-intensity cycling and psychological stress using a modified 
TSST, recorded wearable signals continuously, and sampled cortisol at fixed time 
points across all sessions. We then trained machine learning models to classify five 
states: rest, physical stress, physical recovery, psychological stress, and 
psychological recovery; from 10-minute windows of features.

We hypothesised that: (H1) wearable-derived physiological features alone would 
achieve above-chance classification accuracy across the five stress-related states 
but would show reduced performance for psychological stress and recovery; and (H2) 
integrating salivary cortisol features would significantly improve classification 
performance for psychological stress and psychological recovery, primarily by 
reducing confusion with rest.

\section*{Methods}

\subsection*{Participants}
Six healthy adults (3 female, 3 male; mean age 23.5 years) participated in the 
study. Participants were eligible if they reported no history of cardiovascular, 
endocrine, or psychiatric conditions and were not taking medications known to affect 
autonomic or hormonal function. Eligibility was assessed via a simple health screening 
questionnaire administered by the research team. All participants provided written 
informed consent prior to participation, and the study protocol was approved by the 
Ethics Committee of the Department of Engineering at the University of Cambridge 
(6/9/2018, IONBIKE).

Given the proof-of-concept nature of the study, a convenience sample of six 
participants was recruited. This sample size was chosen to allow intensive 
within-subject measurements across multiple sessions and conditions while remaining 
logistically feasible.

\subsection*{Study Design and Protocol}
\begin{figure*}
\centering
\includegraphics[width=\linewidth]{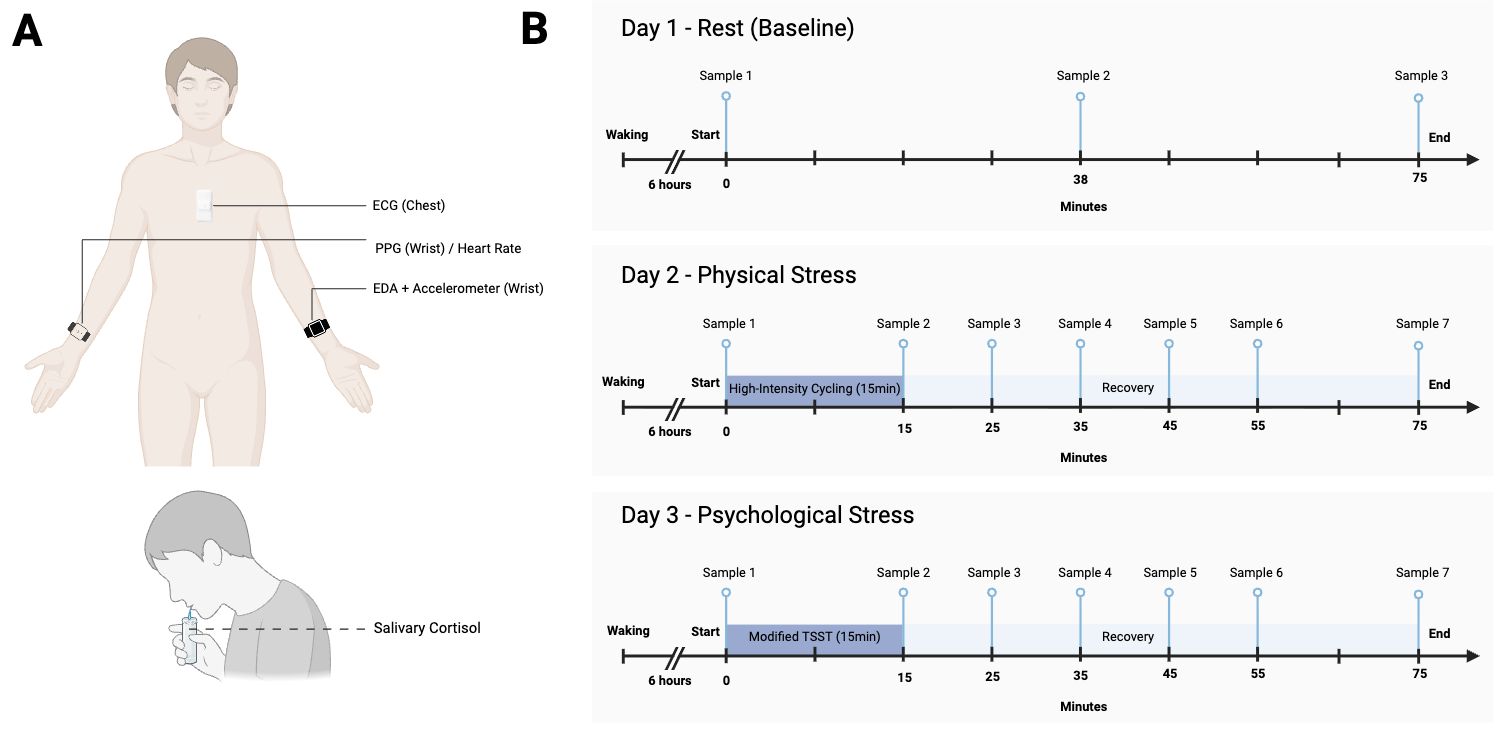}
\caption{Study instrumentation and experimental protocol. (A) Wearable and 
biospecimen measures, including chest electrocardiography (ECG), wrist-based heart 
rate, wrist electrodermal activity (EDA) and accelerometry, and salivary cortisol 
sampling. (B) Protocol timeline for the three experimental sessions: rest 
(baseline), physical stress (high-intensity cycling), and psychological stress 
(modified Trier Social Stress Test). Vertical markers indicate salivary cortisol 
sampling time points.}
\label{fig1}
\end{figure*}

The study employed a within-subject, repeated-measures experimental design 
comprising three sessions, each completed on a separate day: rest (baseline), 
physical stress, and psychological stress. Sessions were scheduled approximately 
six hours after participants' usual wake time (between 12:00 PM and 3:00 PM) to 
minimise variability related to the cortisol awakening response and were separated 
by a washout period of at least 24 hours. Participants were instructed to avoid 
vigorous exercise, alcohol, and caffeine for at least 8 hours prior to each session. 
The order of the three sessions was counterbalanced across participants using a 
Latin square design to minimise potential order effects on cortisol and autonomic 
responses. An overview of the study instrumentation and experimental protocol is 
shown in \Cref{fig1}.

Each session lasted approximately 75 minutes. During the \textbf{rest (baseline) 
session}, participants remained seated in a quiet room and were instructed to relax 
and read neutral magazines or engage in quiet conversation with the researcher. 
During the \textbf{physical stress session}, participants completed 15 minutes of 
high-intensity cycling at a target intensity of 75\% heart rate reserve (HRR), 
followed by a seated recovery period. Heart rate was monitored continuously using 
real-time feedback displayed on the cycling ergometer to ensure that the target 
intensity was maintained throughout. During the \textbf{psychological stress 
session}, participants completed a modified TSST consisting of three phases: a 
preparation phase, a mock job interview, and a mental arithmetic task, followed by 
a seated recovery period. Psychological stress was induced through social-evaluative 
threat and time pressure. Across all sessions, physiological signals were recorded 
continuously, and participants remained seated following the stress tasks to allow 
assessment of recovery dynamics. At the end of each session, participants provided 
a single self-reported stress rating on a 0--10 scale (0 = not stressed, 
10 = extremely stressed).

\subsection*{Wearable Data Collection}
Physiological signals were recorded continuously throughout all experimental 
sessions using wearable sensors to capture autonomic responses associated with 
stress. Cardiac activity was measured using a chest-worn single-lead 
electrocardiography (ECG) sensor (Anne, Sibel Health, USA) positioned on the 
sternum according to manufacturer guidelines. The ECG module samples at 250 Hz and 
provided real-time heart rate monitoring as well as beat-to-beat interval data for 
HRV analysis. Electrodermal activity and wrist movement were recorded using a 
wrist-worn sensor (Empatica E4, Empatica, Italy) worn on the non-dominant wrist, 
which measured skin conductance and three-axis acceleration.

All wearable devices were synchronised prior to each session by establishing 
simultaneous start markers on both sensor outputs. Temporal synchronisation was 
verified post-hoc by visual inspection of rapid transient events (e.g., the onset 
of physical exercise) recorded across both devices. Signal quality was monitored in 
real time during data collection, and any periods with device disconnection or 
movement artifacts were flagged for visual inspection during preprocessing. 
Recordings were acquired continuously across baseline, stress, and recovery periods 
within each session, yielding approximately 4.5 hours of physiological data per 
participant across the three experimental sessions.

\subsection*{Salivary Cortisol Collection}
Salivary cortisol samples were collected during each experimental session at 
predefined time points relative to task onset (\Cref{fig1}). This sampling schedule 
was designed to capture baseline cortisol levels prior to stress exposure, the 
immediate response period during which cortisol begins to rise, and the delayed 
peak and recovery phases characteristic of HPA axis activation following 
psychosocial stress.

Participants were instructed not to eat, drink, smoke, or brush their teeth for 
15 minutes prior to sampling to avoid contamination. Samples were labelled with 
participant ID, session type, and time point, then immediately stored at 
$-20^\circ$C until batch analysis. Cortisol concentrations were quantified using a 
high-sensitivity enzyme-linked immunosorbent assay. Cortisol values are reported 
in micrograms per litre ($\mu$g/L; 1 $\mu$g/dL = 10 $\mu$g/L).

\subsection*{Signal Processing and Feature Extraction}
Physiological and hormonal data streams were temporally aligned using session start 
markers recorded simultaneously on all devices, and all preprocessing and feature 
extraction procedures were predefined and applied consistently across participants 
and sessions.

\subsubsection*{Feature Aggregation} All features were aggregated into nonoverlapping 
10-minute windows to balance temporal resolution with physiological relevance. This 
window length was chosen because autonomic responses to acute stress typically 
stabilise within several minutes, and 10-minute epochs provide sufficient data for 
robust statistical feature extraction while preserving temporal dynamics of stress 
and recovery. For cortisol, which was sampled at discrete time points, features 
were derived by linear interpolation between adjacent samples to estimate cortisol 
levels throughout each session.

Each feature window was labelled as one of five states according to the experimental 
protocol: (1) rest, (2) physical stress, (3) physical recovery, (4) psychological 
stress, or (5) psychological recovery. State labels were assigned based on session 
type and elapsed time relative to task onset. An overview of feature categories is 
provided in \Cref{tab1}.

\begin{table}
\vspace{1.2em}
\caption{Overview of Feature Categories Extracted from Wearable Physiological 
Signals and Salivary Cortisol}
\label{tab1}
\resizebox{\columnwidth}{!}{%
\begin{tabular}{ll}
\hline
\textbf{Modality} & \textbf{Feature Categories} \\
\hline
Heart Rate & Mean HR, HR SD, HR slope \\
HRV (time domain) & SDNN, RMSSD \\
EDA & Tonic, SCR frequency, SCR amplitude \\
Acceleration & Mean magnitude, magnitude SD, z-axis sum \\
Cortisol & Baseline-corrected mean, slope, AUC, raw level, change \\
\hline
\end{tabular}}
\end{table}

\subsection*{Machine Learning Classification and Evaluation}
Two feature sets were constructed: (1) wearable-only model (18 wearable features) 
and (2) multimodal model (18 wearable features + 5 cortisol features). Both models 
used a gradient boosting classifier (XGBoost; XGBClassifier) with pre-specified 
hyperparameters (n\_estimators = 700, eval\_metric = ``mlogloss,'' 
random\_state = 50).

Performance was assessed using leave-one-participant-out cross-validation (LOOCV). 
To ensure subject-independent evaluation and prevent information leakage, data from 
one participant were held out for testing while models were trained on data from the 
remaining participants. All preprocessing steps, including feature standardisation 
and class weighting to account for imbalanced state distributions, were performed 
within each training fold and applied to the corresponding test data.

Model performance was evaluated using overall classification accuracy and class-wise 
precision, recall, and F1-score. Precision was defined as the proportion of samples 
predicted as a given class that were correct, and recall as the proportion of true 
samples of that class that were correctly identified. The F1-score was defined as 
the harmonic mean of precision and recall:
\begin{equation}
F1 = 2 \times \frac{P \times R}{P + R}
\end{equation}
where $P$ denotes precision and $R$ denotes recall. Confusion matrices were examined 
to characterise patterns of misclassification between stress and recovery states.

\section*{Results}

\subsection*{Participant Characteristics}
All six participants completed all three sessions. No adverse events occurred. 
Usable wearable and cortisol data were obtained from all participants across all 
sessions.

\subsection*{Subjective Stress Ratings}
\begin{figure}[!ht]
\centering
\includegraphics[width=1.05\linewidth]{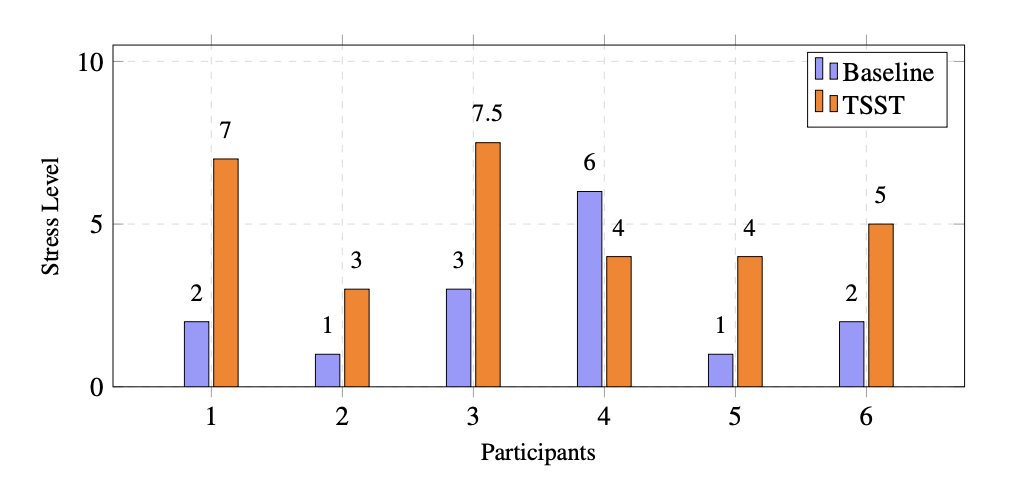}
\caption{Self-reported stress ratings during the TSST and baseline session for 
each of the six participants.}
\label{fig4}
\end{figure}

Participants provided a single self-reported stress rating per session on a 0--10 
scale. As shown in \cref{fig4}, five of six participants reported higher stress 
during the TSST than at baseline (mean baseline = 2.2, mean TSST = 4.8), supporting 
the effectiveness of the stress induction protocol. One participant reported a higher 
baseline rating (6) than TSST rating (4), which may reflect an elevated pre-session 
state due to external circumstances on that day rather than a failure of stress 
induction.

\subsection*{Physiological and Cortisol Responses Across Conditions}
\Cref{fig2} summarises the trajectories of heart rate, HRV (RMSSD), tonic EDA, 
and baseline-corrected salivary cortisol across the rest, physical stress, and 
psychological stress sessions.

\subsubsection*{Physical Stress Session}
As expected, cycling produced pronounced autonomic changes. Heart rate rose from a 
baseline mean of 68.2 bpm to a peak of 176.3 bpm during exercise, then gradually 
declined during recovery. RMSSD decreased from 210 ms at baseline to 18.4 ms during 
exercise, indicating reduced vagal tone, and increased again during recovery. Tonic 
EDA increased from 0.02 S to 6.7 S during exercise, reflecting marked sympathetic 
activation, and decreased towards baseline over the recovery period. Salivary 
cortisol changed little across the session, with no substantial delayed peak, 
indicating limited HPA axis engagement by this physical stressor in this cohort.

\subsubsection*{Psychological Stress Session}
The modified TSST produced qualitatively different dynamics. Heart rate increased 
from 68.2 bpm to 97.3 bpm, representing a moderate autonomic response smaller than 
that seen in the physical stress session. RMSSD decreased from 210 ms to 202 ms, 
and tonic EDA rose to 1.8 S, indicating increased arousal but again with smaller 
magnitude than during cycling. In contrast to the physical session, cortisol 
exhibited a clear delayed response: baseline-corrected levels remained near zero 
during the TSST itself but peaked approximately 40--60 minutes after task onset, 
reaching a maximum of $+$1.18 nmol/L before beginning to decline. This delayed 
cortisol peak aligns with canonical HPA axis responses to social-evaluative stress.

\subsubsection*{Rest Session}
In the rest session, heart rate, HRV, and EDA remained near baseline throughout. 
Cortisol drifted slightly but did not exhibit a systematic rise, consistent with 
the absence of strong stressors.

\begin{figure*} [!ht]
\centering
\includegraphics[width=\linewidth]{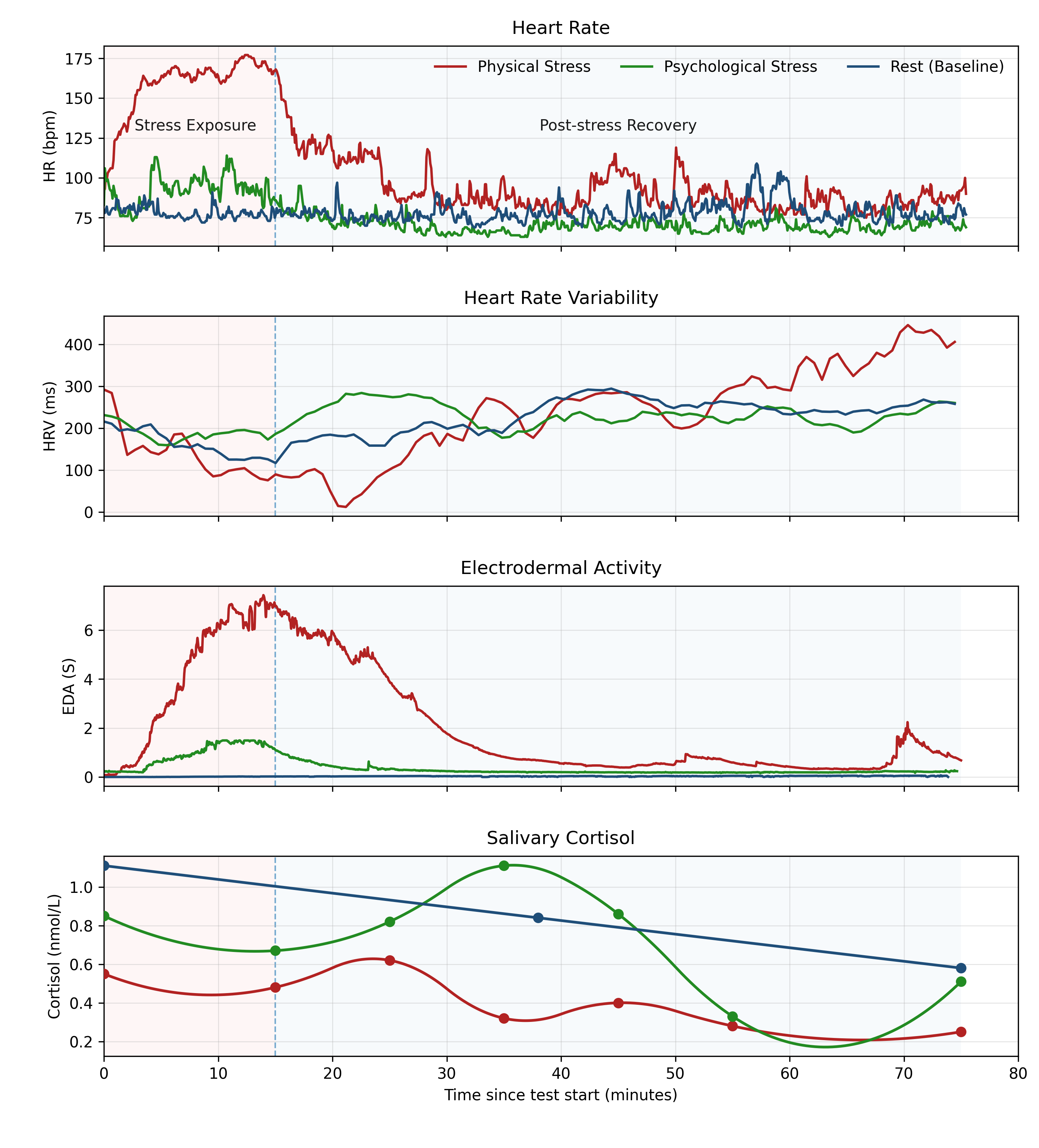}
\caption{Temporal profiles of physiological signals and salivary cortisol during 
rest, physical stress, and psychological stress. From top to bottom: heart rate 
(HR), HRV (RMSSD), EDA tonic skin conductance, and salivary cortisol 
(baseline-corrected). Time 0 marks task onset.}
\label{fig2}
\end{figure*}

\subsection*{Classification Using Wearable Physiological Features Alone}
Using wearable-derived physiological features alone (features from HR, HRV, EDA, 
and acceleration), five-class stress classification achieved an overall accuracy of 
77.8\%. Macro-averaged F1-score was 77.9\%. Performance varied substantially across 
stress states. Physical recovery was identified with the highest reliability (recall 
= 95.8\%), reflecting the distinctive autonomic signature of post-exercise recovery. 
Physical stress also showed strong performance (recall = 83.3\%), consistent with 
the large-magnitude autonomic changes induced by high-intensity exercise. In 
contrast, recall for psychological stress and psychological recovery was markedly 
lower at 50.0\% and 54.2\%, respectively, with frequent misclassification as rest.

The confusion matrix (\Cref{fig3}A) reveals that misclassifications predominantly 
involved psychological states and rest. Psychological stress was misclassified as 
rest in 50\% of samples, and psychological recovery was misclassified as rest in 
42\% of samples (row-normalised proportions). This pattern indicates limited 
separability between psychologically induced stress states and rest when using 
wearable physiological features alone.

\subsection*{Classification with Cortisol Integration}
After incorporating salivary cortisol features, five-class stress classification 
achieved an overall accuracy of 94.4\%. Improvements were most pronounced for 
psychological states: recall for psychological stress increased from 50.0\% to 
83.3\%, and recall for psychological recovery increased from 54.2\% to 87.5\%.

\begin{figure*} [!ht]
\centering
\includegraphics[width=\linewidth]{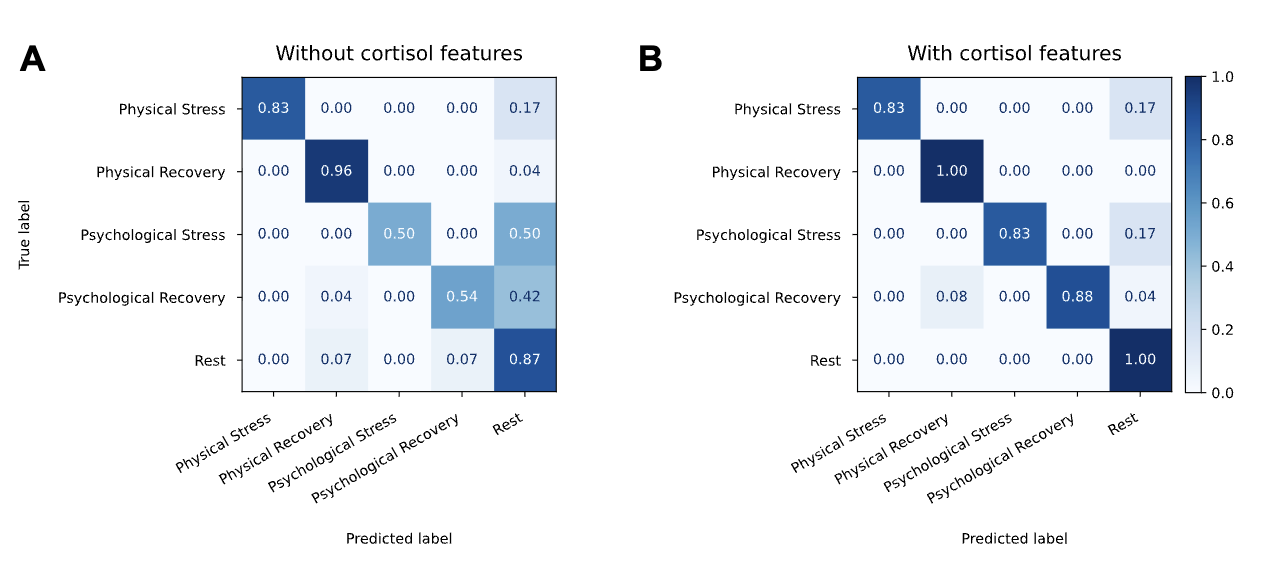}
\caption{Confusion matrices for five-class stress classification (A) using wearable 
physiological features alone and (B) after inclusion of salivary cortisol features. 
Values are row-normalised proportions (each row sums to 1.0). The colour scale 
ranges from 0.0 to 1.0, with darker shading indicating higher proportions.}
\label{fig3}
\end{figure*}

As shown in \Cref{fig3}B, confusion between psychological states and rest was 
substantially reduced. Misclassification of psychological stress as rest decreased 
from 50\% to 17\%, and misclassification of psychological recovery as rest decreased 
from 42\% to 4\% (row-normalised proportions). Recall for rest increased from 86.7\% 
to 100\%, and recall for physical recovery increased from 95.8\% to 100\%. Full 
class-wise performance metrics are reported in \Cref{tab2} and \Cref{tab3}.

\begin{table} [!ht]
\vspace{1.5em}
\caption{Five-Class Stress Classification Performance with Salivary Cortisol 
Integration}
\label{tab2}
\resizebox{\columnwidth}{!}{%
\begin{tabular}{lccc}
\hline
\textbf{Class} & \textbf{Precision (\%)} & \textbf{Recall (\%)} & \textbf{F1 (\%)} \\
\hline
Physical stress & 100.0 & 83.3 & 90.9 \\
Physical recovery & 92.3 & 100.0 & 96.0 \\
Psychological stress & 83.3 & 83.3 & 83.3 \\
Psychological recovery & 100.0 & 87.5 & 93.3 \\
Rest (Baseline) & 93.8 & 100.0 & 96.8 \\
\hline
Overall accuracy & --- & --- & 94.4 \\
Macro average & 96.1 & 88.3 & 89.5 \\
\hline
\end{tabular}}
\end{table}

\begin{table} [!ht]
\vspace{1.3em}
\caption{Five-Class Stress Classification Recall: Wearable-Only Versus 
Multimodal (Wearable + Cortisol) Models}
\label{tab3}
\resizebox{\columnwidth}{!}{%
\begin{tabular}{lccc}
\hline
\textbf{Class} & \textbf{Wearable-Only (\%)} & \textbf{Multimodal (\%)} & 
\textbf{Change (\%)} \\
\hline
Physical stress & 83.3 & 83.3 & 0.0 \\
Physical recovery & 95.8 & 100.0 & $+$4.4 \\
Psychological stress & 50.0 & 83.3 & $+$66.6 \\
Psychological recovery & 54.2 & 87.5 & $+$61.4 \\
\hline
\end{tabular}}
\end{table}

\subsection*{Feature Importance}
Permutation importance analysis identified cortisol slope and mean 
baseline-corrected cortisol as the most informative features for distinguishing 
psychological states from rest. These cortisol features contributed strongly to 
the separation of psychological stress and recovery from low-arousal states, 
reflecting the delayed HPA axis response shown in \Cref{fig2}.

Wearable features, particularly mean heart rate, HRV indices, and tonic EDA, 
remained the dominant contributors for classifying physical stress and recovery. 
The multimodal model thus effectively combined rapid autonomic signatures (for 
identifying physical states) with slower hormonal signatures (for identifying 
psychological states and their recovery).

\section*{Discussion}

\subsection*{Principal Results}
This proof-of-concept study examined whether physical stress, psychological stress, 
and their corresponding recovery states can be distinguished using wearable-derived 
physiological signals and whether integrating salivary cortisol improves 
stress-state classification. Using wearable features alone, a five-class classifier 
achieved an overall accuracy of 77.8\%, with high recall for physical stress (83.3\%) 
and physical recovery (95.8\%), but substantially lower recall for psychological 
stress (50.0\%) and psychological recovery (54.2\%). Psychological states were 
frequently misclassified as rest, mirroring the overlap in autonomic profiles 
between modest psychological stress and low-arousal conditions.

When salivary cortisol features were added, overall accuracy increased to 94.4\%. 
Improvements were most pronounced for psychological states: recall for psychological 
stress rose to 83.3\%, and recall for psychological recovery increased to 87.5\%. 
Confusion between psychological states and rest was markedly reduced. These findings 
indicate that wearable-derived autonomic signals alone are not sufficient to reliably 
differentiate psychological stress from rest and recovery, but that adding endocrine 
information can resolve much of this ambiguity.

\subsection*{Linking Physiology, Cortisol, and Classification}
The classification patterns are consistent with the physiological profiles observed 
across conditions. High-intensity cycling produced large, rapid changes in heart 
rate, HRV, and EDA, generating a distinct autonomic signature that the wearable-only 
model could recognise reliably. By contrast, the TSST induced smaller autonomic 
changes, closer in magnitude to those observed during rest and recovery, making 
psychological stress more difficult to separate from low-arousal states using 
wearable data alone.

At the same time, psychological stress, but not physical stress or rest, elicited 
a delayed cortisol response that peaked approximately 40--60 minutes after task 
onset. Cortisol thus provided a slower, state-specific signal of psychological 
stress exposure and recovery that is not reflected in autonomic features. 
Incorporating cortisol features allowed the model to use both the magnitude and 
timing of HPA axis activation as contextual information, improving its ability to 
label windows whose autonomic characteristics alone were ambiguous.

\subsection*{Implications for Wearable Stress Monitoring}
These findings have important implications for the design and interpretation of 
wearable-based stress metrics. Many current commercial and research systems aggregate 
autonomic signals into a single continuous ``stress score'' without explicitly 
distinguishing whether elevated arousal arises from physical exertion, psychological 
stress, or postexercise recovery. Our results demonstrate that this approach is 
problematic for psychological stress detection: wearable-only models achieved high 
accuracy for physical stress and recovery ($>$83\% recall) but performed only 
slightly above chance for psychological states ($\sim$50\% recall), with half of 
psychological stress episodes misclassified as rest.

This limitation is particularly critical for mental health and stress-aware 
applications, such as digital interventions for anxiety, workplace stress monitoring, 
or early warning systems for stress-related disorders. If psychological stress cannot 
be reliably distinguished from rest and recovery, feedback to users and clinicians 
may be misleading or counterproductive.

The present results show that adding a complementary modality---here, salivary 
cortisol---can substantially improve classification of psychological stress and its 
recovery, increasing recall from 50\% to 83\% for psychological stress and from 
54\% to 88\% for psychological recovery. Several promising directions exist for 
practical implementation. First, emerging wearable cortisol sensors based on sweat 
sampling may enable more frequent hormonal measurements without participant burden. 
Second, proxy markers of HPA axis activity, such as changes in skin temperature or 
multimodal fusion of contextual signals (e.g., location, social context, 
self-reported mood) may approximate the information provided by direct cortisol 
measurement. Third, hybrid approaches combining continuous wearable sensing with 
intermittent biomarker sampling could improve psychological stress detection while 
remaining feasible for real-world deployment.

\subsection*{Limitations}
This work was intentionally designed as a small, tightly controlled proof-of-concept 
study. The sample size was limited to six healthy young adults, which restricts the 
generalisability of the results and leads to relatively wide confidence intervals 
around performance estimates. All stressors were administered under laboratory 
conditions using well-established protocols (cycling and TSST). While this improves 
internal validity, it does not capture the diversity and complexity of stressors 
encountered in daily life.

Cortisol was measured at discrete time points and approximated within 10-minute 
windows using linear interpolation. This approach is adequate to capture the broad 
delayed peak following psychological stress but does not provide fine-grained 
temporal resolution. Furthermore, subjective stress ratings were collected as single 
post-session ratings rather than repeated momentary assessments, which limits their 
temporal resolution and comparability with continuous physiological data.

\subsection*{Future Directions}
Future work should extend these findings in several directions. Larger studies 
with more diverse participants are needed to quantify how robust the observed 
gains from cortisol integration are across individuals, age groups, sex 
differences, and clinical populations. The present sample comprised healthy 
young adults with regular physical activity habits; performance may differ in 
older adults, individuals with chronic stress or anxiety disorders, or those 
with dysregulated HPA axis function. Sample sizes of 30--50 participants would 
provide sufficient statistical power to detect moderate effect sizes and enable 
subgroup analyses.

Real-world ambulatory studies should assess whether similar benefits are 
observed when stressors are less controlled and more heterogeneous. Daily life 
stressors differ from laboratory stress induction in timing, intensity, 
controllability, and social context, all of which may influence autonomic and 
cortisol responses. Combining wearable signals with ecological momentary 
assessment (EMA) to capture self-reported stress, mood, and context would 
strengthen construct validity and enable evaluation of classification performance 
against subjective stress experiences. Cortisol sampling in ambulatory settings 
could be reduced to 2--3 samples per day (e.g., morning, afternoon, evening) to 
balance feasibility with temporal coverage.

Third, alternative or proxy markers of endocrine context should be explored to 
identify practical ways of approximating the information provided by salivary 
cortisol without frequent sampling. Candidates include: (a) other salivary 
biomarkers such as alpha-amylase, which reflects sympathetic-adrenal-medullary 
activity and shows more rapid responses to stress than cortisol; (b) hair 
cortisol, which provides an integrated measure of chronic stress exposure over 
weeks to months and could be used for individual calibration; (c) emerging 
wearable cortisol sensors based on sweat or interstitial fluid sampling, though 
these require validation for stress monitoring; and (d) multimodal fusion of 
contextual signals such as location (e.g., workplace vs.\ home), time of day, 
social interactions, and self-reported mood, which may provide indirect 
information about psychological stress exposure.

Fourth, methodological work is needed to determine how best to fuse autonomic, 
hormonal, and contextual information while respecting the constraints of 
real-time deployment. Promising approaches include: (a) hierarchical models that 
first classify autonomic arousal level and then use cortisol or contextual 
features to disambiguate stress type; (b) time-series models (e.g., recurrent 
neural networks, hidden Markov models) that exploit temporal dynamics of cortisol 
responses, which unfold over 30--60 minutes following psychological stress; (c) 
personalised models that adapt to individual differences in autonomic and hormonal 
reactivity through within-subject calibration; and (d) transfer learning 
approaches that leverage data from multiple individuals to improve generalisation 
to new users with limited calibration data.

Finally, clinical validation studies are needed to assess whether improved 
differentiation of psychological stress translates to clinically meaningful 
outcomes. Potential applications include: (a) monitoring stress burden in 
individuals with anxiety or mood disorders to inform treatment decisions; (b) 
early warning systems that detect prolonged or frequent psychological stress 
episodes as risk indicators for stress-related health problems; (c) real-time 
stress management interventions that deliver coping strategies (e.g., breathing 
exercises, cognitive reappraisal prompts) selectively during detected 
psychological stress; and (d) workplace stress monitoring to identify high-stress 
periods or environments and guide organisational interventions.

\section*{Conclusion}
This proof-of-concept study demonstrates that wearable physiological signals alone 
lack sufficient contextual specificity to reliably differentiate psychological stress 
from rest and recovery, even though they perform well for physical stress and 
physical recovery. Integrating salivary cortisol, a marker of HPA axis activation, 
substantially improved classification of psychological stress and psychological 
recovery and reduced misclassification of these states as rest. These findings 
highlight an important limitation of wearable-only stress monitoring and support 
the development of multimodal approaches that combine autonomic and endocrine 
information for more accurate stress interpretation. Larger and more ecologically 
valid studies are now required to confirm these results and to identify practical 
alternatives to repeated salivary cortisol sampling for real-world stress-aware 
systems.

\begin{acknowledgements}
O. Kaya and M. V. Alban-Paccha conceptualised and designed the study and wrote the manuscript draft. O. Kaya performed the data acquisition, analysis, and interpretation of the data. N. Athanassopoulou and G. G. Malliaras supervised the study and acquired funding. All authors read and contributed to the final manuscript. The authors thank Amos Folarin, Senior Software Development Group Leader at the Biomedical Research Centre, King's College London, for providing additional sensors; Till-Jonas Gerngroß, Head of Business Development at Cerascreen, for facilitating the cortisol analysis and supporting the biochemical validation; Dr. Onur Parlak for early advice on the methodological design, particularly the saliva collection methods; and Andrea Hongn, doctoral researcher at the Instituto de Ingeniería Biomédica, for guidance on multimodal physiological data analysis. The authors declare no conflicts of interest.
\end{acknowledgements}

\section*{Bibliography}

\bibliography{references}

\end{document}